\documentclass[journal=jacsat,manuscript=article]{achemso}
\usepackage{siunitx}
\usepackage{graphicx} 
\usepackage[utf8]{inputenc}
\usepackage[T1]{fontenc}
\usepackage{amsmath}
\usepackage{amsfonts}
\usepackage{amssymb}
\usepackage{array}
\usepackage[version=4]{mhchem} 
\usepackage{braket}
\usepackage{bm}
\usepackage{xr}
\usepackage{xspace}
\usepackage{multirow}
\usepackage{hyperref}
\usepackage{cleveref}
\usepackage{titlesec}

\usepackage{xspace} 
\usepackage{caption}
\usepackage{subcaption}
\usepackage{tikz}
\usetikzlibrary{arrows, arrows.meta}
\usepackage{comment}
\usepackage{newtxtext}
\usepackage{mathtools}
\setkeys{acs}{articletitle = true}

\DeclareUnicodeCharacter{0301}{\'{e}}

\usepackage{xr-hyper}
\makeatletter
\newcommand*{\addFileDependency}[1]{
 \typeout{(#1)}
 \@addtofilelist{#1}
 \IfFileExists{#1}{}{\typeout{No file #1.}}
}
\makeatother

\newcommand*{\wfq}{$\omega$FQ\xspace}
\newcommand*{\wfqfmu}{$\omega$FQF$\mu$\xspace}

\newcommand{\abs}[1]{\left|{#1}\right|}
\newcommand*{\de}{\mathop{}\!\mathrm{d}}
\newcommand*{\imm}{\mathrm{i}}

\newcommand*{\tqq}{\mathbf{T}^{qq}}
\newcommand*{\tqmu}{\mathbf{T}^{q\mu}}
\newcommand*{\tmuq}{\mathbf{T}^{\mu q}}
\newcommand*{\tmumu}{\mathbf{T}^{\mu\mu}}

\newcommand*{\sm}{SI\xspace}

\author{Piero Lafiosca}
\affiliation{Scuola Normale Superiore, Piazza dei Cavalieri 7, 56126 Pisa, Italy}
\author{Luca Nicoli}
\affiliation{Scuola Normale Superiore, Piazza dei Cavalieri 7, 56126 Pisa, Italy}
\author{Silvio Pipolo}
\affiliation{UCCS Unit\'e de Catalyse et Chimie du Solide, Universit\'e de Lille, Universit\'e d’Artois UMR 8181, F-59000, Lille, France}
\author{Stefano Corni}
\affiliation{Dipartimento di Scienze Chimiche, Universit{\`a} di Padova, via Marzolo 1, 35131, Padova, Italy}
\alsoaffiliation{Istituto di Nanoscienze del Consiglio Nazionale delle Ricerche CNR-NANO, via Campi 213/A, 41125, Modena, Italy}
\author{Tommaso Giovannini}
\affiliation{Department of Physics, University of Rome Tor Vergata, Via della Ricerca Scientifica 1, 00133, Rome, Italy}
\email{tommaso.giovannini@uniroma2.it}
\author{Chiara Cappelli}
\affiliation{Scuola Normale Superiore, Piazza dei Cavalieri 7, 56126 Pisa, Italy}
\email{chiara.cappelli@sns.it}

\title[]
  {Real-Time Formulation of Atomistic Electromagnetic Models for Plasmonics}


\begin{document}

\begin{abstract}
Investigating nanoplasmonics using time-dependent approaches permits shedding light on the dynamic optical properties of plasmonic structures, which are intrinsically connected with their potential applications in photochemistry and photoreactivity. This work proposes a real-time extension of our recently developed fully atomistic approaches $\omega$FQ and $\omega$FQF$\mu$. These methods successfully reproduce quantum size effects in metal nanoparticles, including plasmon shifts for both simple and $d$-metals, even below the quantum size limit. Also, thanks to their atomistic nature and the phenomenological inclusion of quantum tunneling effects, they can effectively describe the optical response of subnanometer junctions. By incorporating real-time dynamics, the approach provides an efficient framework for studying the time-dependent optical behavior of metal nanostructures, including the decoherence of plasmon excitations. 
\end{abstract}


\newpage

\section{Introduction}

Under the action of external radiation, the conductive electrons of plasmonic nanomaterials, such as noble metal nanoparticles, can be excited, giving rise to the so-called localized surface plasmons (LSP).\cite{pitarke2006theory,giannini2011plasmonic,halas2011plasmons} Such plasmons are excited at the plasmon resonance frequency (PRF), which generally shows a high tunability as a function of the size, shape, and chemical composition of the nanostructure.\cite{kelly2003optical,liz2006tailoring,ringe2010unraveling,langer2019present} Such a property, together with the capability of plasmonic materials to absorb a huge amount of energy and the huge enhancement of the electric near field, makes plasmonic materials based on noble metals a unique platform for a plethora of technological applications, ranging from sensing\cite{willets2007localized,zhang2013chemical,jiang2015distinguishing,chiang2016conformational,langer2019present,benz2016single,yang2020sub,liu2017single} to photocatalysis.\cite{yuan2022earth,ezendam2022hybrid,baldi2023pulsed,herran2022tailoring,yuan2022plasmonic,baumberg2022picocavities,li2021recent,gargiulo2019optical,mukherjee2013hot,clavero2014plasmon}

In this context, an in-depth understanding of the dynamics of the optical response of plasmonic structures is vital for elucidating the mechanisms underlying the photophysics and photochemistry of such systems.\cite{robatjazi2020plasmon,zhan2020recent,dong2023plasmonic} From the experimental point of view, this can be achieved by resorting to time-resolved spectroscopies.\cite{cortes2022experimental,sa2024dynamics} To rationalize the experimental findings, theoretical approaches able to describe the time evolution of the optical response need to be exploited.\cite{rossi2020hot,liu2017relaxation,sanchez2022plasmon,wu2023molecular} The most common theoretical methods generally rely on solving classical Maxwell's Equations or purely quantum mechanical (QM) methods. Classical approaches, such as the Boundary Element Method (BEM)\cite{myroshnychenko2008modelling,hohenester2012mnpbem,de2002retarded} or the  Finite-Difference Time-Domain (FDTD) method,\cite{hao2007efficient,sehmi2017optimizing} are powerful tools for modeling nanoscale optical systems, however, they typically overlook tunneling effects and the atomistic details that are essential for an accurate description of structures defined at the subnanometer levels,\cite{teperik2013quantum,zhu2016quantum,urbieta2018atomic,savage2012revealing,marinica2012quantum,esteban2012bridging,esteban2015classical,campos2019plasmonic,scholl2013observation,barbry2015atomistic} such as nanojunctions or picocavities.\cite{baumberg2022picocavities} On the other hand, QM-based approaches, headed by real-time time-dependent density functional theory (RT-TDDFT) and TDDFT,\cite{domenis2023time,rossi2020hot,herring2023recent,kuisma2015localized,ding2014quantum,senanayake2019real,baseggio2015new,baseggio2016photoabsorption,sinha2017classical} provide a first-principle description of the nanostructure response by retaining the atomistic nature of the system. However, they are computationally prohibitive for large nanosystems and are generally exploited to describe systems composed at most of a few thousand atoms.

To overcome the drawbacks of classical and QM-based approaches, we have recently developed two methods, namely frequency-dependent fluctuating charges ($\omega$FQ)\cite{giovannini2019classical,bonatti2020frontiers,giovannini2020graphene,bonatti2022silico,zanotto2023strain,yamada2021classical,huang2024time} and frequency-dependent fluctuating charges and fluctuating dipoles ($\omega$FQF$\mu$).\cite{giovannini2022we,nicoli2023fully} Such methods describe the nanostructure at the full atomistic level and model its optical response by integrating the Drude conduction theory, classical electrodynamics, quantum mechanical tunneling, and interband effects. In particular, in \wfq, each atom of the nanosystem is endowed with a charge that dynamically responds to the external field as regulated by the Drude conduction model.\cite{bade1957drude,giovannini2019classical} In this way, intraband effects are incorporated into the model. To describe $d$-metal nanostructures, we have extended \wfq by introducing \wfqfmu, where each atom is additionally endowed with a complex polarizability, modeling $d$ shell polarizability and interband effects.\cite{giovannini2022we} In both approaches, we also introduce a phenomenological description of quantum tunneling effects regulating the charge exchange between neighbor atoms.\cite{esteban2012bridging,esteban2015classical} Both \wfq and \wfqfmu can reproduce reference \emph{ab initio} and experimental data, including plasmon shifts for both simple and noble metals, and atomistically defined nanojunctions, even below the quantum size limit.\cite{giovannini2019classical,giovannini2022we} In other words, they feature accuracy comparable to QM-based approaches, at a much lower computational cost.

In this work, we extend the $\omega$FQ and $\omega$FQF$\mu$ models to the real-time (RT) regime, enabling the simulation of the time evolution of plasmonic excitations as driven by light pulses of arbitrary profiles. The resulting RT-\wfq and RT-\wfqfmu approaches permit the study of the dynamics of the optical response of metal nanostructures, including the decoherence of plasmon excitations. The approaches can describe the response dynamics under short impulses (in the as/fs timescale) and continuous waves (CW) using the same theoretical framework. This is particularly relevant since CW irradiation is often exploited in plasmonic catalysis.\cite{yuan2022earth,robatjazi2020plasmon} We validate the novel approaches by comparing our results with reference \emph{ab initio} TDDFT data for sodium and silver nanostructures. Our method thus offers a comprehensive and efficient tool for studying the dynamics of optical response of metal nanostructures, bridging the gap between classical and QM descriptions, and paving the way for advanced applications in nanophotonics.

The manuscript is organized as follows. In the next section, we report the theoretical derivation of RT-\wfq and RT-\wfqfmu. The methods are then applied to simulate the time evolution of the optical response of a sodium dimer, displaying a single-atom junction, and a silver icosahedral nanoparticle. The summary and conclusions end the manuscript.

\section{Theoretical Model}

In this section, we first briefly recall the theoretical foundations of the fully atomistic \wfq and \wfqfmu approaches for nanoplasmonics. We then present and discuss their extension to the time domain.

\subsection{\texorpdfstring{\wfq}{wfq} and \texorpdfstring{\wfqfmu}{wfqfmu} models}

$\omega$FQ is a fully atomistic, classical model that endows each atom of the nanostructure with a frequency-dependent charge.\cite{giovannini2019classical} When an external electric oscillating field is applied to the system, the charge exchange between the atoms is governed by the Drude conduction model. Such an interaction exponentially vanishes by phenomenologically incorporating quantum tunneling mechanisms limiting the charge transfer among the nearest neighboring atoms.\cite{giovannini2019classical} By considering a monochromatic field oscillating at frequency $\omega$,  $\omega$FQ charges $q_{i}(\omega)$ are computed by solving the following equation:\cite{giovannini2019classical} 
\begin{align}
-i \omega q_i(\omega) & = \frac{2n_0\tau}{1-i\omega\tau}\sum^N_{j} \left[ 1 - f(l_{ij}) \right]\frac{\mathcal{A}_{ij}}{l_{ij}} (\phi_j^{el}-\phi_i^{el}) \nonumber \\
& = \frac{2n_0\tau}{1-i\omega\tau} \sum^N_j K_{ij} (\phi_j^{el}-\phi_i^{el})
\label{eq:wfq}
\end{align}
where $N$ is the number of atoms, $\tau$ is the friction time, $n_0$ is the numerical density of the material, $\mathcal{A}_{ij}$ is the effective area connecting $i$-th and $j$-th atoms, and $l_{ij}$ is their distance. $\phi^{el}_i$ is the electrochemical potential on each atom regulating the dynamical polarization of the system. This takes into account the interactions between the atoms and their interaction with the external electric field. Finally, $f(l_{ij})$ is a Fermi-like function mimicking quantum tunneling:\cite{giovannini2019classical}
\begin{equation}
    f(l_{ij}) = \frac{1}{1+\exp\left[-d \left(\frac{l_{ij}}{s\cdot l^{0}_{ij}} - 1\right)\right]}
\label{eq:fermi}
\end{equation}
where $l^0_{ij}$ is the atom-atom equilibrium distance, whereas $d$ and $s$ determine the sharpness and the center of the damping function, respectively. The parameters entering \cref{eq:wfq} have a microscopic physical meaning and can be extracted from the experimental permittivity or determined by comparing computed results with reference \textit{ab initio} calculations.\cite{giovannini2019classical,giovannini2020graphene} 

\Cref{eq:wfq} can be recast in a compact way as the following linear system:\cite{lafiosca2021going}
\begin{equation}\label{eq:wfq-key}
\left[\mathbf{A}^{qq}-z_q(\omega)\mathbf{I}_N\right]\mathbf{q}(\omega) = - \mathbf{f}^q(\omega)
\end{equation}
where $\mathbf{I}_N$ is the $N \times N$ identity matrix, while $\mathbf{A}^{qq}$, $z_q(\omega)$ and $\mathbf{f}_q (\omega)$ read:\cite{lafiosca2021going} 
\begin{equation}
\begin{aligned}
\mathbf{A}^{qq} & = (\mathbf{K}-\mathbf{P})\tqq \\
z_q(\omega) & = \frac{1}{2n_0}\left(-\omega^2-\imm\omega\frac{1}{\tau}\right) \\
\mathbf{f}^q(\omega) & =  (\mathbf{K}-\mathbf{P})\mathbf{V}^{ext}(\omega)
\end{aligned}
\label{eq:A_zq_bq}
\end{equation}
where $\mathbf{T}^{qq}$ is the charge-charge interaction kernel\cite{giovannini2019classical,mayer2007formulation} and $\mathbf{V}^{ext}(\omega)$ is the potential associated with the external field evaluated at the position of each atom. The matrix $\mathbf{P}$ takes the following form:\cite{lafiosca2021going}
\begin{equation}
P_{ij} = \sum_k K_{ik}\delta_{ij}
\end{equation}
$\omega$FQ is a fully atomistic approach that models the optical response of nanostructure materials by accounting for intraband mechanisms only, \emph{via} the Drude conduction model. Therefore, the approach cannot describe metals featuring interband transitions, such as noble metal nanostructures.\cite{pinchuk2004optical,pinchuk2004influence,balamurugan2005evidence,liebsch1993surface,santiago2020efficiency} To model these systems, we have recently developed $\omega$FQF$\mu$,\cite{giovannini2022we} which extends \wfq by assigning to each atom of the nanosystem an additional source of polarization, i.e., an atomic polarizability and an associated induced dipole $\bm{\mu}_i$. The electric potential produced by the dipoles is included in \cref{eq:wfq} as an external potential acting on the charges in the electrochemical potential $\phi^{el}$. The dipoles $\boldsymbol{\mu}_{i}$ are calculated by solving the following set of linear equations:\cite{giovannini2022we}
\begin{equation}
\bm{\mu}_i = \alpha^\mathrm{IB} (\omega) \left(\mathbf{E}^{q}_i + \mathbf{E}^{\mu}_i + \mathbf{E}_i^{ext} \right)
\label{eq:wfmu}
\end{equation}
where $\mathbf{E}^{q}$ and $\mathbf{E}^{\mu}$ are the electric fields generated by the charge and the other dipole moments, respectively, while $\mathbf{E}^{ext}$ is the external electric field. $\alpha^\mathrm{IB}(\omega)$ is the atomic complex polarizability describing interband transitions, which can be easily obtained by extracting interband contributions from the experimental permittivity function.\cite{giovannini2022we} 

The equations defining charges and dipoles responses can be coupled together resulting in the following linear system:\cite{giovannini2022we} 
\begin{equation}\label{eq:wfqfmu-key}
\left[ 
\left(\begin{array}{cc}
\mathbf{A}^{qq}  & \mathbf{A}^{q\mu} \\ 
\tmuq & \tmumu
\end{array}\right) 
-
\left(
\begin{array}{cc}
z_q(\omega) \mathbf{I}_N & \mathbf{0} \\ 
\mathbf{0} & z_\mu(\omega)\mathbf{I}_{3N}
\end{array}
\right)
\right]
\left(\begin{array}{c}
\mathbf{q}(\omega) \\
\bm{\mu}(\omega)
\end{array}\right)
=
\left(\begin{array}{c}
-\mathbf{f}^q(\omega) \\
\mathbf{f}^\mu(\omega)
\end{array}\right)
\end{equation}
where $\tmuq$ and $\tmumu$ are the dipole-charge and dipole-dipole interaction kernels,\cite{mayer2007formulation,giovannini2019fqfmu,giovannini2022we} while $\mathbf{A}^{q\mu}$, $z_\mu(\omega)$ and $\mathbf{f}^\mu (\omega)$ are defined as:\cite{giovannini2022we}
\begin{equation}
\begin{aligned}
\mathbf{A}^{q\mu} & = (\mathbf{K} - \mathbf{P}) \tqmu \\
z_\mu(\omega) & = -\frac{1}{\alpha^\mathrm{IB}(\omega)} \\
\mathbf{f}^\mu(\omega) & = \mathbf{E}^{ext}(\omega)
\end{aligned}
\end{equation}
where $\tqmu=\left[\tmuq\right]^\dagger$ is the charge-dipole interaction kernel.\cite{mayer2007formulation,giovannini2019fqfmu,giovannini2022we} 

\subsection{Real-Time Extension of \texorpdfstring{\wfq}{wfq} and \texorpdfstring{\wfqfmu}{wFQFu} models}\label{sec:theory_rt}

In this section, we extend both \wfq and \wfqfmu to the time domain by proposing their real-time extension. Let us first focus on \wfq. Explicating the definition of $z_q(\omega)$ in \cref{eq:wfq-key}, we obtain:
\begin{equation}\label{eq:wfq-w}
\left(-\frac{\omega^2}{2n_0}-\frac{\imm\omega}{2n_0\tau}\right)\mathbf{q}(\omega) = \mathbf{A}^{qq} \mathbf{q}(\omega) + \mathbf{f}^q(\omega)
\end{equation}
By considering the following definition of the Fourier transform
\begin{equation}\label{eq:ft}
\mathcal{F}[f](\omega) =\int_{-\infty}^\infty\de t\, e^{\imm\omega t} f(t),\quad \mathcal{F}^{-1}[f](t) = \frac{1}{2\pi}\int_{-\infty}^\infty\de \omega\, e^{-\imm\omega t} f(\omega) 
\end{equation}
we can rewrite \cref{eq:wfq-w} in the time domain by applying $\mathcal{F}^{-1}$ to both sides 
\begin{equation}\label{eq:wfq-rt}
\frac{1}{2n_0}\ddot{\mathbf{q}}(t)+\frac{1}{2n_0\tau}\dot{\mathbf{q}}(t) = \mathbf{A}^{qq} \mathbf{q}(t) + \mathbf{f}^q(t)
\end{equation}
where we exploited the following properties of the Fourier transform:
\begin{equation}\label{eq:ft-props}
\begin{aligned}
\dot{\mathbf{q}}(t) & = \frac{1}{2\pi} \int_{-\infty}^\infty \de\omega\, e^{-\imm\omega t}(-\imm\omega)\mathbf{q}(\omega) \\
\ddot{\mathbf{q}}(t) & = \frac{1}{2\pi} \int_{-\infty}^\infty \de\omega\, e^{-\imm\omega t}(-\omega^2)\mathbf{q}(\omega)
\end{aligned}
\end{equation}
\Cref{eq:wfq-rt} represents the dynamical evolution of a set of coupled forced damped oscillators with the same mass $\frac{1}{2n_0}$ and the same damping parameter $\frac{1}{\tau}$, under the time-dependent force $\mathbf{f}_q(t)$ that depends on the electric potential generated on each charge at the time $t$. \Cref{eq:wfq-rt} defines the RT-\wfq approach.

By moving to \wfqfmu, \cref{eq:wfqfmu-key} can be rewritten explicating $z_q(\omega)$ and $z_\mu(\omega)$ as follows:
\begin{align}
\left(-\frac{\omega^2}{2n_0}-\frac{\imm\omega}{2n_0\tau}\right)\mathbf{q}(\omega) & = \mathbf{A}^{qq} \mathbf{q}(\omega) + \mathbf{A}^{q\mu} \bm{\mu}(\omega) + \mathbf{f}^q(\omega)\label{eq:wfqfmu-charges}\\
\bm{\mu}(\omega) & = \alpha^\mathrm{IB}(\omega)\left[-\tmuq\mathbf{q}(\omega) - \tmumu\bm{\mu}(\omega) + \mathbf{f}^\mu(\omega)\right] \label{eq:wfqfmu-dipoles}
\end{align}
Before applying the inverse Fourier transform, it is convenient to rewrite \cref{eq:wfqfmu-dipoles} to avoid the convolution in the time domain in terms of interband polarizability. To this end, we approximate the interband polarizability $\alpha^\mathrm{IB}(\omega)$ as a sum of Drude-Lorentz oscillators (DL), similarly to what has been proposed in FDTD\cite{hao2007efficient,sehmi2017optimizing} and BEM\cite{dall2020real} frameworks:
\begin{equation}\label{eq:alpha-ib}
\alpha^\mathrm{IB}(\omega) \approx \alpha^\mathrm{IB}_{fit}(\omega) 
= \sum_{p}^M \frac{A_p}{\omega_p^2-\omega^2-\imm\omega\gamma_p}
\end{equation}
where $M$ is the total number of DL oscillators, each of them defined in terms of the parameters $A_p,\omega_p$, and $\gamma_p$. We can split \cref{eq:wfqfmu-dipoles} by partitioning $\bm{\mu}(\omega)$ as DL-dependent terms $\bm{\mu}_p(\omega)$ such that:
\begin{align}
\bm{\mu}(\omega) & = \sum_{p}^M\bm{\mu}_p(\omega) \\
\bm{\mu}_p(\omega) & = \frac{A_p}{\omega_p^2-\omega^2-\imm\omega\gamma_p}\left[-\tmuq\mathbf{q}(\omega) - \tmumu\bm{\mu}(\omega) + \mathbf{f}^\mu(\omega)\right]
\end{align}
At this point, the original \wfqfmu problem in \cref{eq:wfqfmu-charges} and \cref{eq:wfqfmu-dipoles} can be recast in $M+1$ equations in the frequency domain, i.e.: 
\begin{equation}\label{eq:wfqfmu-split}
\begin{aligned}
\left(-\frac{\omega^2}{2n_0}-\frac{\imm\omega}{2n_0\tau}\right)\mathbf{q}(\omega) = & \mathbf{A}^{qq} \mathbf{q}(\omega)+ \mathbf{A}^{q\mu} \bm{\mu}(\omega) + \mathbf{f}^q(\omega) \\
\frac{\omega_p^2-\omega^2-\imm\omega\gamma_p}{A_p}\bm{\mu}_p(\omega) = & - \tmuq\mathbf{q}(\omega) - \tmumu\bm{\mu}(\omega) + \mathbf{f}^\mu(\omega)
\end{aligned}
\end{equation}
By applying $\mathcal{F}^{-1}$ Fourier transform to both charges and dipoles expressions in \cref{eq:wfqfmu-split}, we finally obtain the $M + 1$ differential equations defining the RT-\wfqfmu model:
\begin{align}
\frac{1}{2n_0}\ddot{\mathbf{q}}(t)+\frac{1}{2n_0\tau}\dot{\mathbf{q}}(t) & = \mathbf{A}^{qq}\mathbf{q}(t)+ \mathbf{A}^{q\mu}\bm{\mu}(t) + \mathbf{f}^q(t) \label{eq:wfqfmu-rt-charges} \\
\frac{1}{A_p}\ddot{\bm{\mu}}_p(t) + \frac{\gamma_p}{A_p}\dot{\bm{\mu}}_p(t) & =  - \tmuq\mathbf{q}(t) - \tmumu\bm{\mu}(t) + \mathbf{f}^\mu (t) - \frac{\omega_p^2}{A_p}\bm{\mu}_p(t)  \label{eq:wfqfmu-rt-dipoles}
\end{align}
where we have exploited again the properties recalled in \cref{eq:ft-props}.

\section{Numerical propagation scheme}

To propagate RT-\wfq and RT-\wfqfmu in time, we exploit a second-order velocity Verlet algorithm in the presence of a friction term.\cite{vanden2006second} Note that such a procedure has been previously exploited to deal with similar problems in the context of continuum approaches.\cite{dall2020real} 
Let us consider a general coordinate system $\mathbf{x}(t)$ described by the following differential equation:
\begin{equation}
    \ddot{\mathbf{x}}(t) = \mathbf{F}(t) - \gamma\dot{\mathbf{x}}(t)
\end{equation}
where $\mathbf{F}$ is the mass-normalized force acting on each coordinate and $\gamma$ is the friction coefficient. By using a fixed step $\Delta t$, the second-order velocity Verlet algorithm scheme expresses $\mathbf{x}(t)$ and the associated velocity $\dot{\mathbf{x}}(t)$ as:
\begin{equation}\label{eq:second-order}
\begin{aligned}
\mathbf{x}(t+\Delta t) & = \mathbf{x}(t) + \Delta t\left(1-\gamma\frac{\Delta t}{2}\right)\dot{\mathbf{x}}(t) + \frac{\Delta t^2}{2}\mathbf{F}(t) \\
\dot{\mathbf{x}}(t+\Delta t) & = \left(1-\gamma\Delta t + \gamma^2\frac{\Delta t^2}{2}\right)\dot{\mathbf{x}}(t) + \frac{\Delta t}{2}\left(1 - \gamma\Delta t\right)\mathbf{F}(t) + \frac{\Delta t}{2}\mathbf{F}(t+\Delta t)
\end{aligned}
\end{equation}
Eq. \ref{eq:second-order} can be specified for RT-\wfq and RT-\wfqfmu by defining the mass-normalized time-dependent force acting on charges and dipoles from \cref{eq:wfqfmu-rt-charges,eq:wfqfmu-rt-dipoles}:
\begin{align}
\mathbf{F}^q(t) & = 2 n_0 \left [ \mathbf{A}^{qq} \mathbf{q}(t) + \mathbf{A}^{q\mu} \bm{\mu}(t) + \mathbf{f}^{q}(t) \right] \label{eq:force_q}\\
\mathbf{F}^\mu_p(t) & = A_p \left[- \tmuq\mathbf{q}(t) - \tmumu\bm{\mu}(t) + \mathbf{f}^{\mu}(t) \right] - \omega_p^2\bm{\mu}_p(t) \label{eq:force_mu}
\end{align}
Note that for RT-\wfq, only the forces acting on charges are calculated by discarding $\mathbf{A}^{q\mu} \bm{\mu}(t)$ in \cref{eq:force_q}. 

The second-order velocity Verlet algorithm scheme in \cref{eq:second-order} can finally be specified for the time-propagation of RT-\wfq as follows:
\begin{align}
\mathbf{q}(t+\Delta t) & =  \mathbf{q}(t) + \Delta t\left(1 -\frac{\Delta t}{2\tau}\right)\dot{\mathbf{q}}(t) + \frac{\Delta t^2}{2} \mathbf{F}^q(t) \label{eq:q_t}\\
\dot{\mathbf{q}}(t+\Delta t) & = \left(1-\frac{\Delta t}{\tau}+\frac{\Delta t^2}{2\tau^2}\right)\dot{\mathbf{q}}(t)+\frac{\Delta t}{2}\left(1 - \frac{\Delta t}{\tau}\right)\mathbf{F}^q(t) + \frac{\Delta t}{2}\mathbf{F}^q(t+\Delta t) \label{eq:qpunto_t} 
\end{align}
In RT-\wfqfmu, the dipoles' time-propagation can be obtained similarly as:
\begin{align}
\bm{\mu}_p(t+\Delta t) & =  \bm{\mu}_p(t) + \Delta t\left(1-\gamma_p\frac{\Delta t}{2}\right)\dot{\bm{\mu}}_p(t) + \frac{\Delta t^2}{2}\mathbf{F}^\mu(t) \label{eq:mu_t} \\
\dot{\bm{\mu}}_p(t+\Delta t) & =  \left(1-\gamma_p\Delta t+\gamma_p^2\frac{\Delta t^2}{2}\right)\dot{\bm{\mu}}_p(t) + \frac{\Delta t}{2}\left(1 - \gamma_p\Delta t\right)\mathbf{F}^\mu(t) + \frac{\Delta t}{2}\mathbf{F}^\mu(t+\Delta t) \label{eq:mupunto_t}
\end{align}
\section{Computational Details}

RT-\wfq and RT-\wfqfmu equations are solved by using a stand-alone Fortran 95 package. The numerical integration of charges and dipoles is performed by propagating \cref{eq:q_t,eq:qpunto_t,eq:mu_t,eq:mupunto_t} from $t=0$ and by assuming zero initial conditions, which naively correspond to the stationary solution in the absence of external field. \wfq and \wfqfmu parameters for Na and Ag nanostructures are recovered from Ref. \citenum{giovannini2019classical} and \citenum{giovannini2022we}, respectively. 

For RT-\wfqfmu, the interband polarizability $\alpha^\mathrm{IB}$ in \cref{eq:alpha-ib} is fitted to experimental data by resorting to the Basin-hopping stochastic global optimization algorithm.\cite{wales1997global} The Ag interband polarizability (see Ref. \citenum{giovannini2022we} for its definition) is fitted with a different number of DL oscillators ($M = 4, 5, 6, 7$) by minimizing the residual error: 
\begin{equation}\label{eq:residual}
\mathrm{res} = \abs{\int_{\omega_{min}}^{\omega_{max}}\de\omega \, \alpha^\mathrm{IB}(\omega)-\alpha^\mathrm{IB}_{fit}(\omega)}
\end{equation}
where $\alpha^\mathrm{IB}_{fit}$ is the fitted permittivity that depends on the parameters $A_p,\omega_p$, and $\gamma_p$. For each value of $M$, the optimal parameters are obtained by selecting the best $\alpha^\mathrm{IB}_{fit}$ from 100 Basin-hopping simulations. Each Basin-hopping simulation is started by setting random values of $\omega_p$ extracted in the range between $\omega_{min}=1$ eV and $\omega_{max} = 7$ eV, $A_p$ and $\gamma_p$ equal to 1 for each $p=1,\dots,M$. The fitting procedure is implemented in Python by resorting to the \verb|lmfit| package\cite{newville_2015_11813}, which exploits the Basin-hopping algorithm as implemented in the \verb|scipy| library.\cite{2020SciPy-NMeth} Note that for $M=6,7$ the fitting procedure returns some values of $\omega_p\approx 0$ which can potentially generate numerical issues in the time propagation. However, these values can be safely discarded because $\alpha^\mathrm{IB}_{fit}$ is correctly reproduced in the experimental frequency region of interest (vide infra). Three sets of parameters with $M=4,5,6$ are finally generated. The numerical values of $A_p,\omega_p$, and $\gamma_p$ are reported in table S1 in the Supporting Information -- \sm. 

\section{Numerical Applications}

In this section, RT-\wfq and RT-\wfqfmu approaches are first validated by comparing the absorption spectra as calculated by Fourier-transform the time-dependent responses or by using \wfq and \wfqfmu defined in the frequency domain. RT-\wfq and RT-\wfqfmu robustness is further demonstrated by simulating the time-resolved spectral signal of a Sodium dimer characterized by an atomistically defined junction and the dynamical response of an Icosahedral Silver NP and comparing our results with reference \emph{ab initio} data. 

\subsection{Validation of the propagation scheme}\label{sec:validation}

\begin{figure}[!htbp]
    \centering
    \includegraphics[width=.47\textwidth]{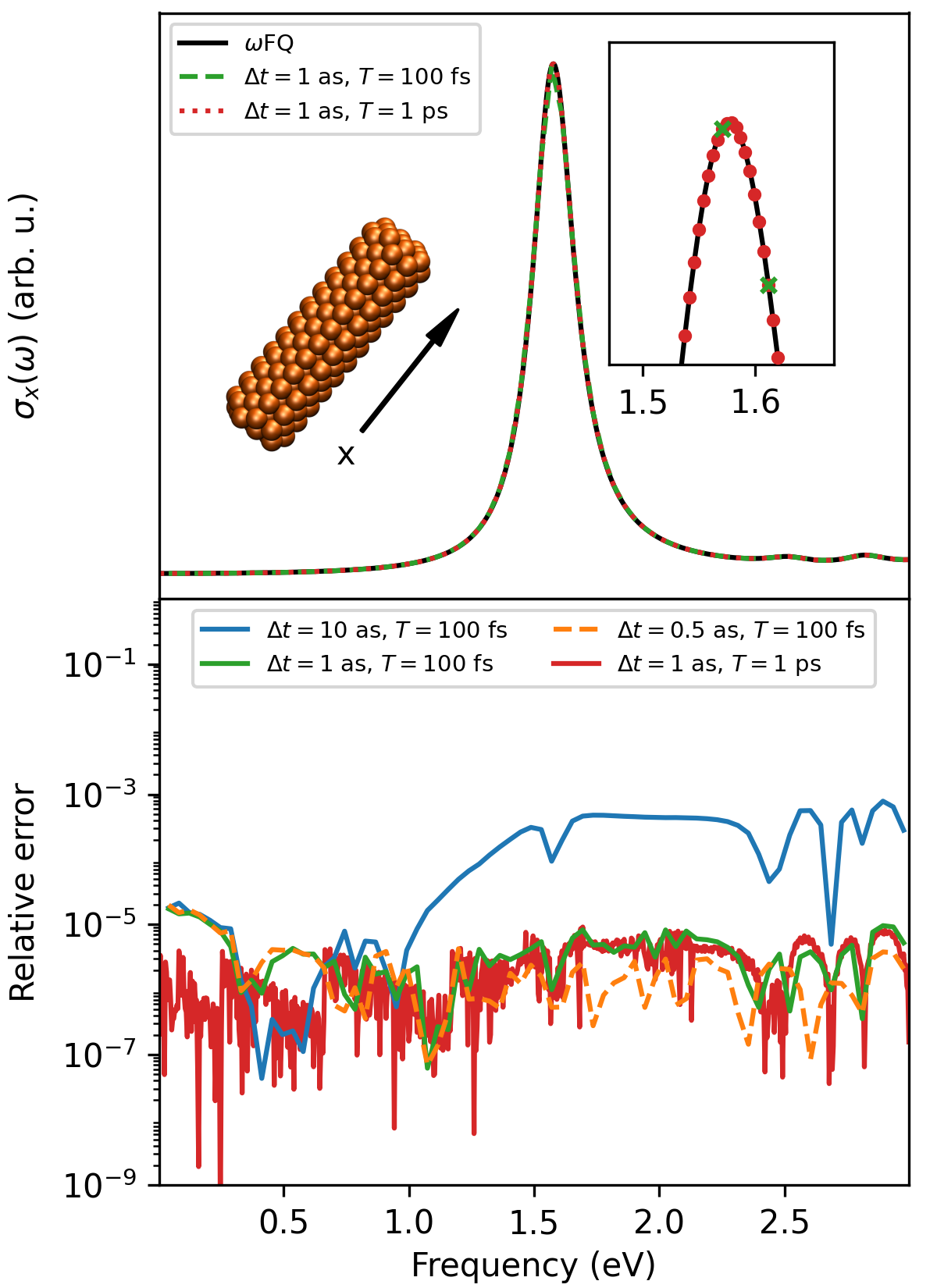}
    \includegraphics[width=.47\textwidth]{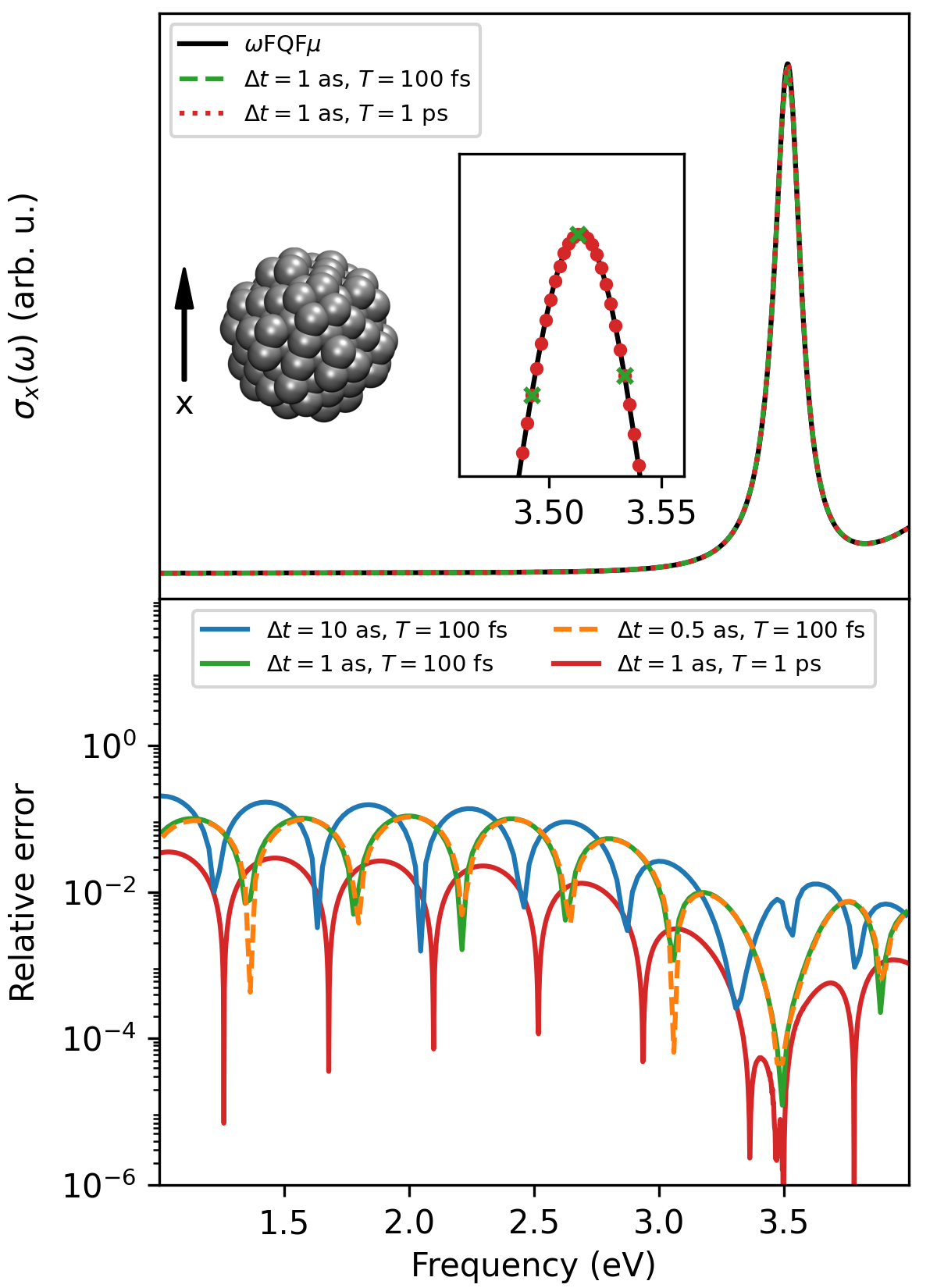}
    \caption{(left) RT-\wfq and \wfq absorption cross section ($\sigma_x$) of a \ce{Na261} nanorod as polarized along the $x$ axis. RT-\wfq results are obtained by varying the parameters of the time evolution ($\Delta t$ and $T$) under the action of the external pulse in \cref{eq:kick-pulse}. Relative RT-\wfq-\wfq errors are reported in the bottom panel. (right) RT-\wfqfmu and \wfqfmu absorption cross section ($\sigma_x$) of a \ce{Ag164} spherical NP as polarized along the $x$ axis. RT-\wfqfmu results are obtained by varying the parameters of the time evolution ($\Delta t$ and $T$) under the action of the external pulse in \cref{eq:kick-pulse}. Relative RT-\wfq-\wfq errors are reported in the bottom panel.}
    \label{fig:kick-validation}
\end{figure}

RT-\wfq and RT-\wfqfmu are validated by computing the optical response of a Sodium cylindrical nanorod (\ce{Na261}, length 4.968 nm, diameter 1.242 nm, described at the RT-\wfq level), and a silver spherical NP (\ce{Ag164}, diameter 1.634 nm, described at the RT-\wfqfmu level). The considered nanostructures are graphically depicted in \cref{fig:kick-validation}. For both systems, we apply an external electric field described as the following Gaussian-type kick pulse linearly polarized along the $\hat{\mathbf{x}}$ axis:
\begin{equation}\label{eq:kick-pulse}
\mathbf{E}^{ext}_{\Delta t}(t) = E_0 \exp{\left[-\frac{(t-t_0)^2}{2(\sigma\Delta t)^2} \right] }\hat{\mathbf{x}},\quad 0<t<T
\end{equation}
where $t_0=5$ fs, $\sigma=0.1$, while $\Delta t$ and $T$ are the time-step and the total evolution time exploited in the time propagation. The Gaussian width depends on $\Delta t$ in order to have a uniform sampling by varying the time step.

The longitudinal absorption cross-section $\sigma_k$ along the axis $k=x,y,z$ is computed as:
\begin{equation}
    \sigma_k(\omega) = \frac{4\pi\omega}{c} \mathrm{Im}[\alpha_{kk}(\omega)]
    \label{eq:cross_section}
\end{equation}
where $\omega$ is the frequency, $c$ is the speed of light and $\alpha_{kk}(\omega)$ is the frequency-dependent polarizability element, which is obtained from the time-propagation dynamics as follows: 
\begin{align}\label{eq:polar-fourier}
\alpha^{\text{RT-}\omega\text{FQ}}_{kk}(\omega) & = \frac{\int\de t \left[\sum_i^N q_i(t) r_{i,k} \right]e^{\imm\omega t} }{\int\de t E^{ext}_k(t)e^{\imm\omega t} } \nonumber \\
\alpha^{\text{RT-}\omega\text{FQF}\mu}_{kk}(\omega) & = \frac{\int\de t \left[\sum_i^N q_i(t) r_{i,k} + \mu_{i,k}(t)\right]e^{\imm\omega t} }{\int\de t E^{ext}_k(t)e^{\imm\omega t} }
\end{align}
where $q_i$ is the charge of the $i$-th atom, whereas $r_{i,k}$, $\mu_{i,k}$ and $E_k$ are the $k = {x,y,z}$ Cartesian components of the position and the electric dipole of atom $i$, and the external electric field.
The Fourier transform in \cref{eq:polar-fourier} is numerically computed by resorting to the Fast Fourier Transform as implemented in the \verb|scipy| package.\cite{2020SciPy-NMeth}

RT-\wfq and RT-\wfqfmu longitudinal absorption cross sections of the \ce{Na261} nanorod and \ce{Ag164} spherical NP are reported in \cref{fig:kick-validation} (top panel), where they are compared to the corresponding $\sigma_x(\omega)$ calculated by exploiting the reference \wfq and \wfqfmu models, respectively. The relative error on $\sigma_x(\omega)$ obtained from RT-$\omega$FQ(F$\mu$) with respect to $\omega$FQ(F$\mu$) is reported in the bottom panel of \cref{fig:kick-validation}. In the case of (RT-)\wfqfmu calculations, the fitted interband permittivity $\alpha^\mathrm{IB}_{fit}$ obtained with $M=6$ (see table S1 in the \sm) is used for both RT-\wfqfmu and \wfqfmu, thus ensuring the same optical response of the system in the two cases. 

To validate our integration scheme, the absorption cross-section is calculated by varying the integration time step $\Delta t$ (0.5, 1, 10 as) and the total simulation time $T$ (100 fs, 1 ps). By first inspecting \cref{fig:kick-validation}, we note that the numerical results obtained from real-time simulations are qualitatively consistent with the reference values calculated in the frequency domain, perfectly reproducing all the features of the absorption spectrum for both \wfq and \wfqfmu. This is valid for both $T=100$ fs and $T=1$ ps, for which the most accurate results are obtained as a higher density of data points is acquired (see inset of \cref{fig:kick-validation}, top panel). 

For a more quantitative analysis, we investigate the relative error obtained for RT-\wfq and RT-\wfqfmu by taking their frequency-domain counterparts as a reference. Such an analysis is performed for the same input frequencies in both the real-time and frequency-domain calculations. The numerical results are depicted in \cref{fig:kick-validation}, bottom panel, by varying the numerical parameters of the integration scheme ($\Delta t$, $T$) to quantify their impact on the numerical stability of the simulation. By first focusing on RT-\wfq (left), we observe that a large $\Delta t$ ($10$ as; $T = 100$ fs) is generally associated with the largest errors, in particular in correspondence of the maxima in the absorption spectrum, i.e., for frequencies $> 1.5$ eV. Differently, all calculations performed by using a smaller time-step ($0.5,1$ as; $T=100$ fs), are associated with a substantial reduction of relative errors for all frequencies and are overall consistent with each other. Interestingly, elongating the total time of the simulation does not significantly affect the numerical errors in the region of the spectrum characterized by absorption peaks. 

The relative errors computed for RT-\wfqfmu are inherently higher than those reported for RT-\wfq by almost 2 orders of magnitude on average. In this case, reducing the time step of the integration does not substantially reduce the relative errors. The most accurate results are obtained by increasing the precision of the Fast Fourier Transform, i.e., by elongating the time simulation of the real-time propagation ($T = 1$ ps), which is associated with an increase in the number of sampled frequencies. This is particularly evident in the significant regions of the spectrum (frequency $>$ 3 eV), where the relative errors are comparable to those reported by RT-\wfq. The relatively large errors can thus be attributed to limitations in numerical precision.

The discussed analysis validates the accuracy of our implementation and the exploited second-order procedure. In all the following calculations, the time step $\Delta t = 1$ ps is exploited because it guarantees the best compromise between computational cost and accuracy.

\subsection{Sodium NP dimer: single-atom junction}

\begin{figure}[!htbp]
    \centering
    \includegraphics[width=0.43\textwidth]{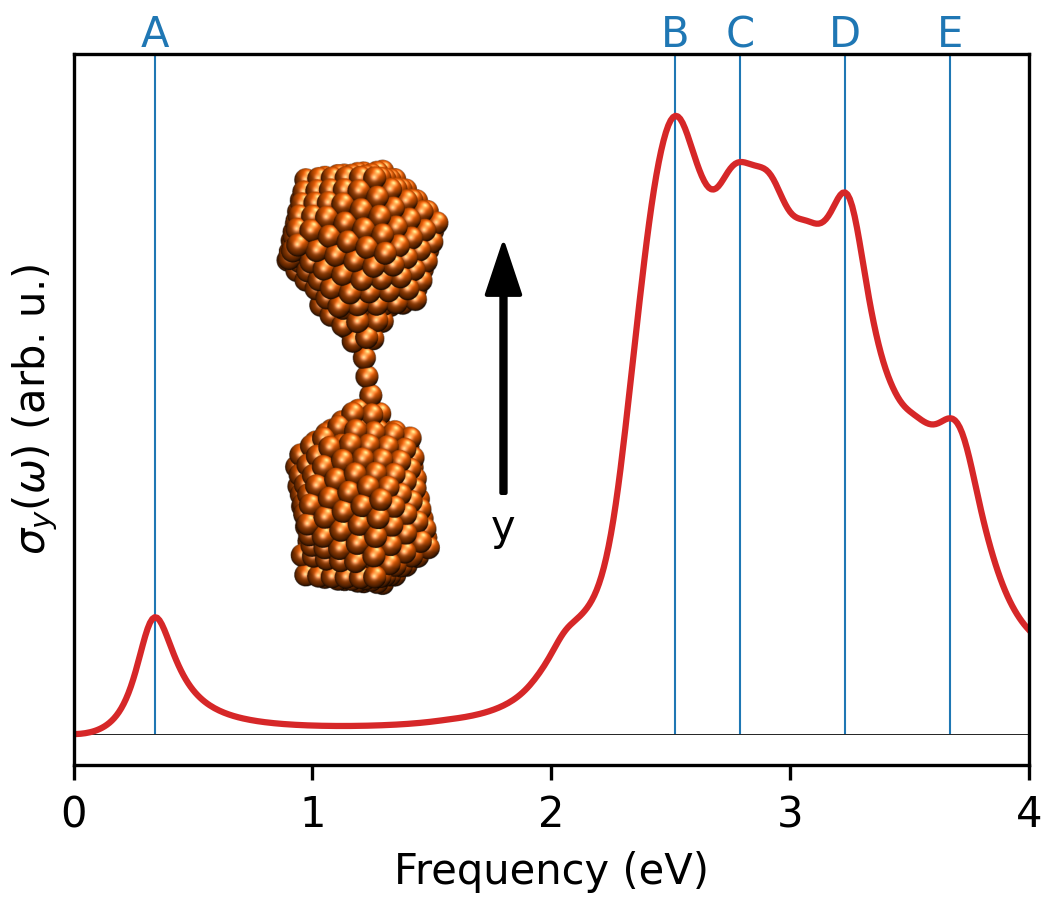} \\
    \includegraphics[width=0.43\textwidth]{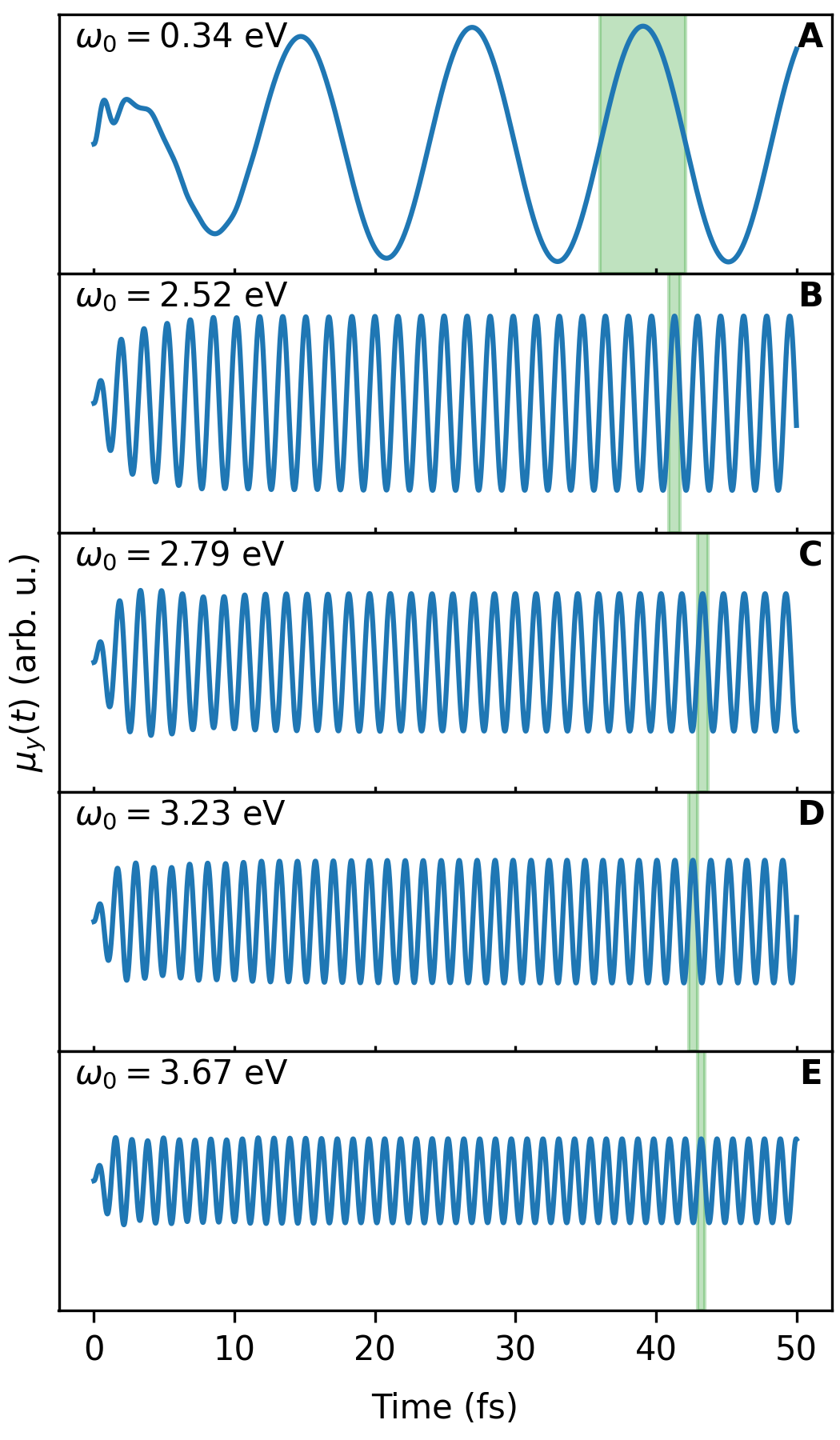}
    \caption{(top) \wfq absorption cross section of Na$_{380}$ dimer as polarized along the $y$ axis. (bottom) RT-\wfq total dipole moment induced in the $y$ direction as a function of time for A, B, C, D, and E peaks highlighted in the top panel.}
    \label{fig:na380-retracting}
\end{figure}

As a first test case to analyze the performance of RT-\wfq, we consider two Na$_{380}$ icosahedral NPs that are connected by a monoatomic junction (see inset in \cref{fig:na380-retracting}). Such structure is extracted from a dynamical simulation of the retraction process of two fused \ce{Na380} NPs, which has been 
studied at the full \emph{ab-initio} level by Marchesin et al.\cite{marchesin2016plasmonic} and at the \wfq level in Ref. \citenum{giovannini2019classical}. The breaking process gradually occurs: the monoatomic junction first arises, and then the dimer dissociates as the distance increases. This study focuses on the monoatomic junction because we have recently shown that tunneling effects are essential to reproduce the \emph{ab initio} reference data.\cite{giovannini2019classical,marchesin2016plasmonic} As such, this system represents a perfect test case to show the capabilities of the newly developed RT-\wfq for studying the dynamics of the optical response of atomistically defined NPs. 

In fact, the simulated absorption cross-section along the longitudinal axis of the NP dimer ($y$) at the \wfq level is characterized by two main bands.\cite{giovannini2019classical} Such plasmon peaks are associated with two peculiar charge-transfer (CT) plasmonic excitations. The band occurring at $0.34$ eV (see A peak in \cref{fig:na380-retracting}) is characterized by a dipolar plasmon in the whole dimer structure. For this reason, it is generally named Charge-Transfer Plasmon (CTP).\cite{liu2013plasmon,perez2011optical,duan2012nanoplasmonics} The second band dominates the spectrum and is especially broad ($2-4$ eV), resulting from the convolution of many absorption peaks. This behavior is commonly identified in most NP dimers.\cite{rossi2015quantized,marchesin2016plasmonic,urbieta2018atomic,varas2016quantum,zhang2014ab} The associated plasmon shows an overall multipolar character (generally dipolar character by looking at the single nanoparticles), and it is generally called CTP'. The large inhomogeneous broadening reported for the CTP' band is due to transitions with diﬀerent nodal structure at the atomic scale, but corresponding to plasmons of similar nature (see B, C, D, and E peaks in \cref{fig:na380-retracting}), similarly to what has been reported for other stretched sodium NPs.\cite{rossi2015quantized,giovannini2019classical} 

To study the plasmon dynamics in the time domain, we resort to RT-\wfq. We exploit an external cosinusoidal electric field along the dimer axis ($y$): 
\begin{equation}
    \mathbf{E}^{ext}(t) = E_0 \cos (\omega_0 t) \hat{\mathbf{y}}\label{eq:e_ext_na}
\end{equation}
Differently from \cref{sec:validation}, the used functional form resembles a continuous wave (CW) illumination of the nanostructure. This demonstrates the flexibility of our approach to describing different excitations, which can better represent the experimental conditions.\cite{yuan2022earth,robatjazi2020plasmon} The real-time simulation is carried out by using $E_0=10^{-6}$ au ($\approx 50\mu V$/\AA ), $\Delta t = 1$ as, and $T = 50$ fs. In \cref{fig:na380-retracting}, the time propagation of the total dipole moment along the dimer axis $\mu_y$ is plotted by setting $\omega_0$ (see \cref{eq:e_ext_na}) in resonance with A, B, C, D, and E absorption frequencies. The numerical values are normalized with respect to the dipole moments computed at resonance with A absorption, for which the largest $\mu_y$ is reported. 
By looking at \cref{fig:na380-retracting}, we note that the stationary solution is reached for all external frequencies after about 20 fs. In such a time regime, the total dipole moment oscillates in phase with the external field at the forcing frequency, which increases moving from the A to the E band. Remarkably, the relaxation time is similar for each external frequency, since it is intrinsically related to the scattering time assigned to sodium atoms ($\tau$, see \cref{eq:wfq}). Furthermore, the relative amplitudes of the dipole moments associated with the diverse plasmon excitations are directly connected with the imaginary part of the polarizability of the system, which in turn is related to the absorption cross section reported in \cref{fig:na380-retracting} through \cref{eq:cross_section}. This is why the largest amplitudes are reported for $\omega_0$ in resonance with the A peak, and the relative amplitudes decrease by increasing the external frequency, following the trends highlighted by the absorption spectrum.

\begin{figure}[!htbp]
    \centering
    \includegraphics[width=0.7\textwidth]{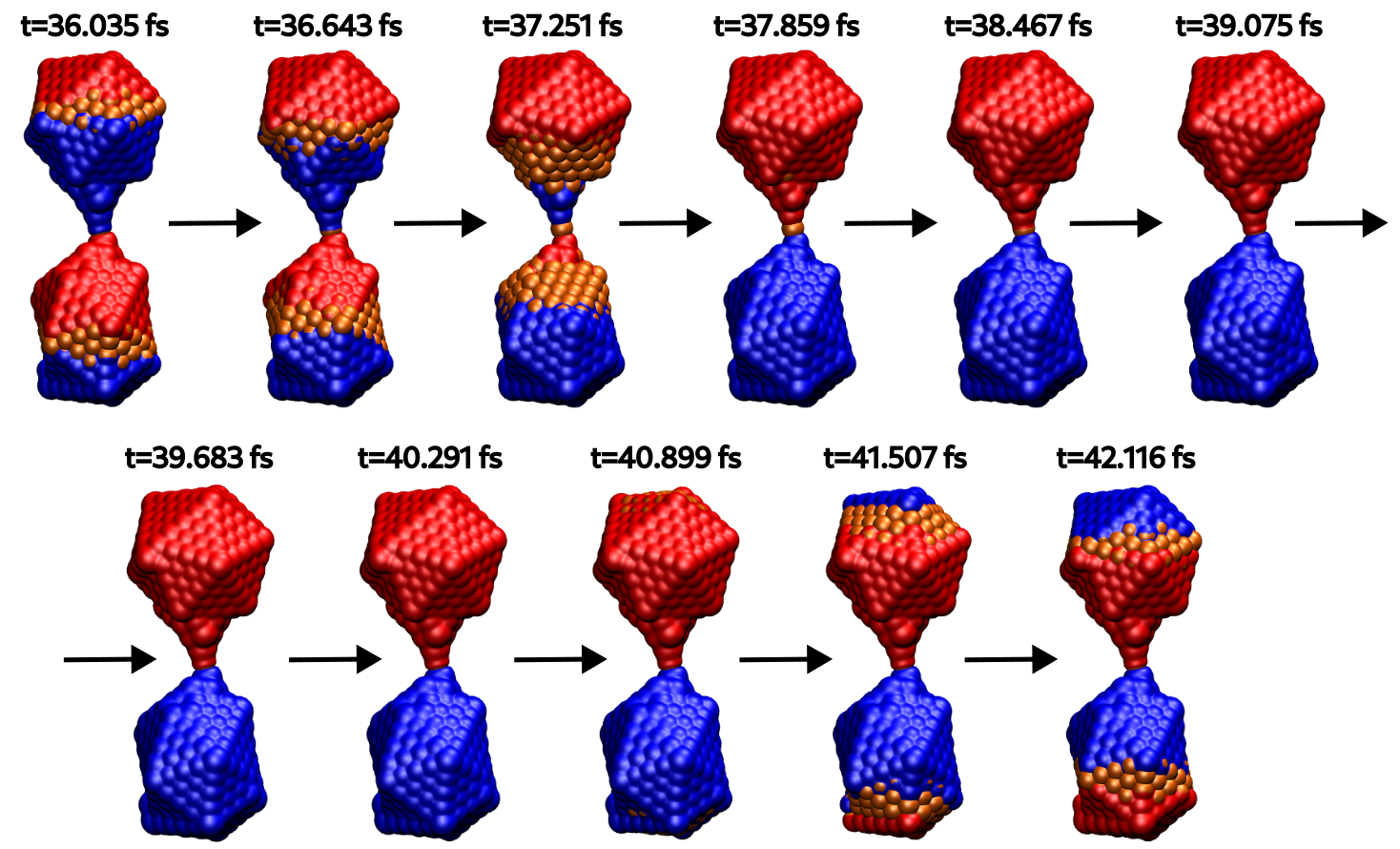}
    \caption{RT-\wfq time evolution of the plasmon excitation of \ce{Na380} dimer excited using a monochromatic field oscillating at 0.34 eV (A peak in \cref{fig:na380-retracting}, top panel). The plasmon oscillation is reported for the time region highlighted in green in \cref{fig:na380-retracting}, bottom panel. The isovalue is set to 0.001 au.}
    \label{fig:na380-dynamics}
\end{figure}

Our real-time model also allows for graphically investigating the dynamical oscillation of the plasmonic excitation. A similar study has also been performed at the \emph{ab initio} level as derived from TDDFT calculations, assuming a periodic oscillation equal to $2\pi/\omega_0$.\cite{marchesin2016plasmonic} RT-\wfq gives direct access to this analysis, allowing an in-depth investigation of the dynamics of the plasmon. In this case, we focus on plasmonic dynamics once the system enters the stationary condition. Such a time evolution is depicted in \cref{fig:na380-dynamics} for the CTP peak at $0.34$ eV (all the other time evolution are graphically depicted in figs. S2 to S5 in the \sm). As can be noticed, the single-atom junction behaves as an accumulation point for electron conduction, limiting the charge transfer between the two structures. Such a structure is associated with a substantial reduction of the electron current across the single-atom junction at both the \wfq and TDDFT levels.\cite{marchesin2016plasmonic,giovannini2019classical} We finally remark that the plasmon dynamics in \cref{fig:na380-dynamics} qualitatively reproduces the TDDFT behavior\cite{marchesin2016plasmonic} at a much lower computational cost.




\subsection{Icosahedral Ag NP}

\begin{figure}[!htbp]
    \centering
    \includegraphics[width=.52\textwidth]{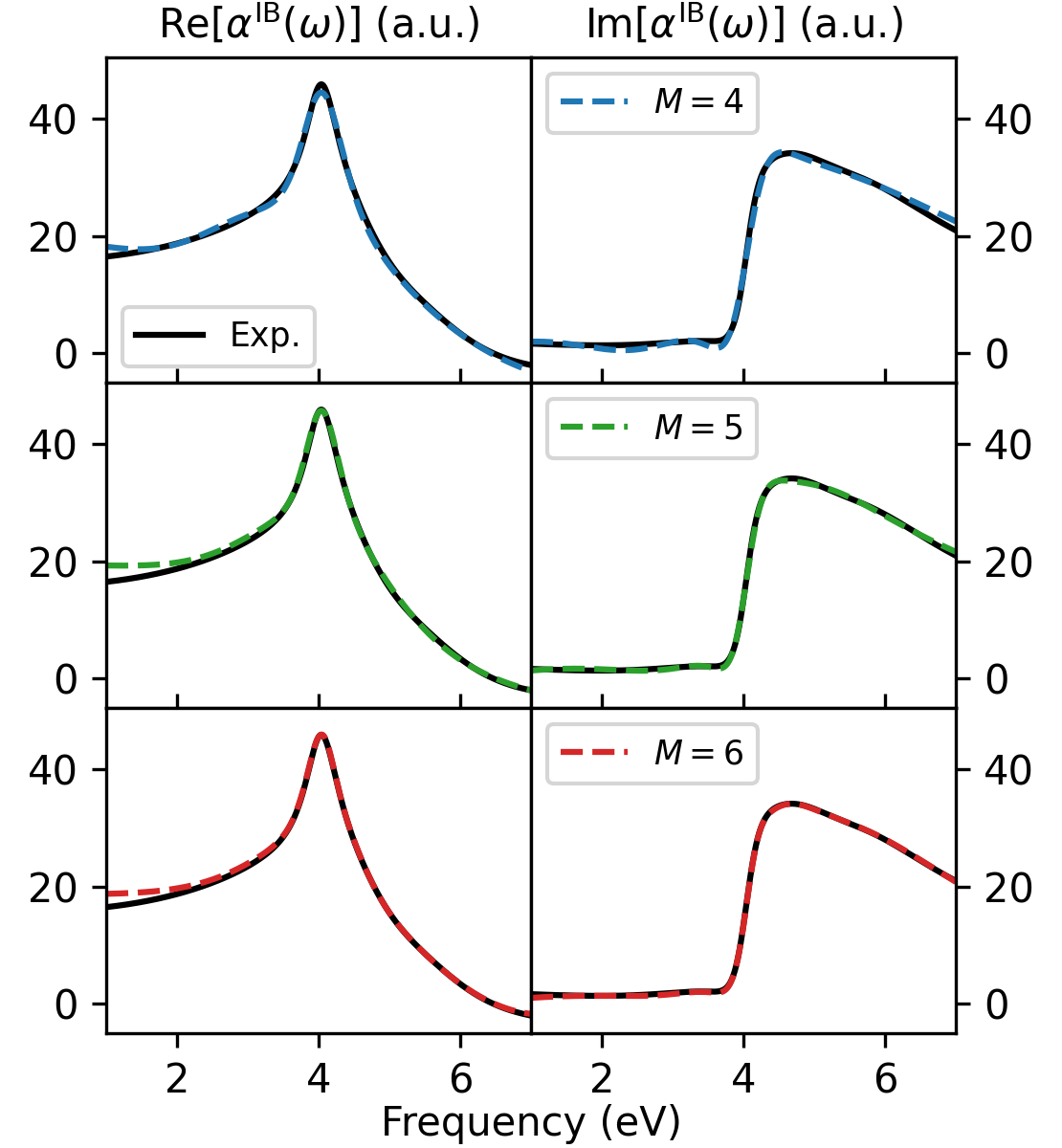}
    \includegraphics[width=.42\textwidth]{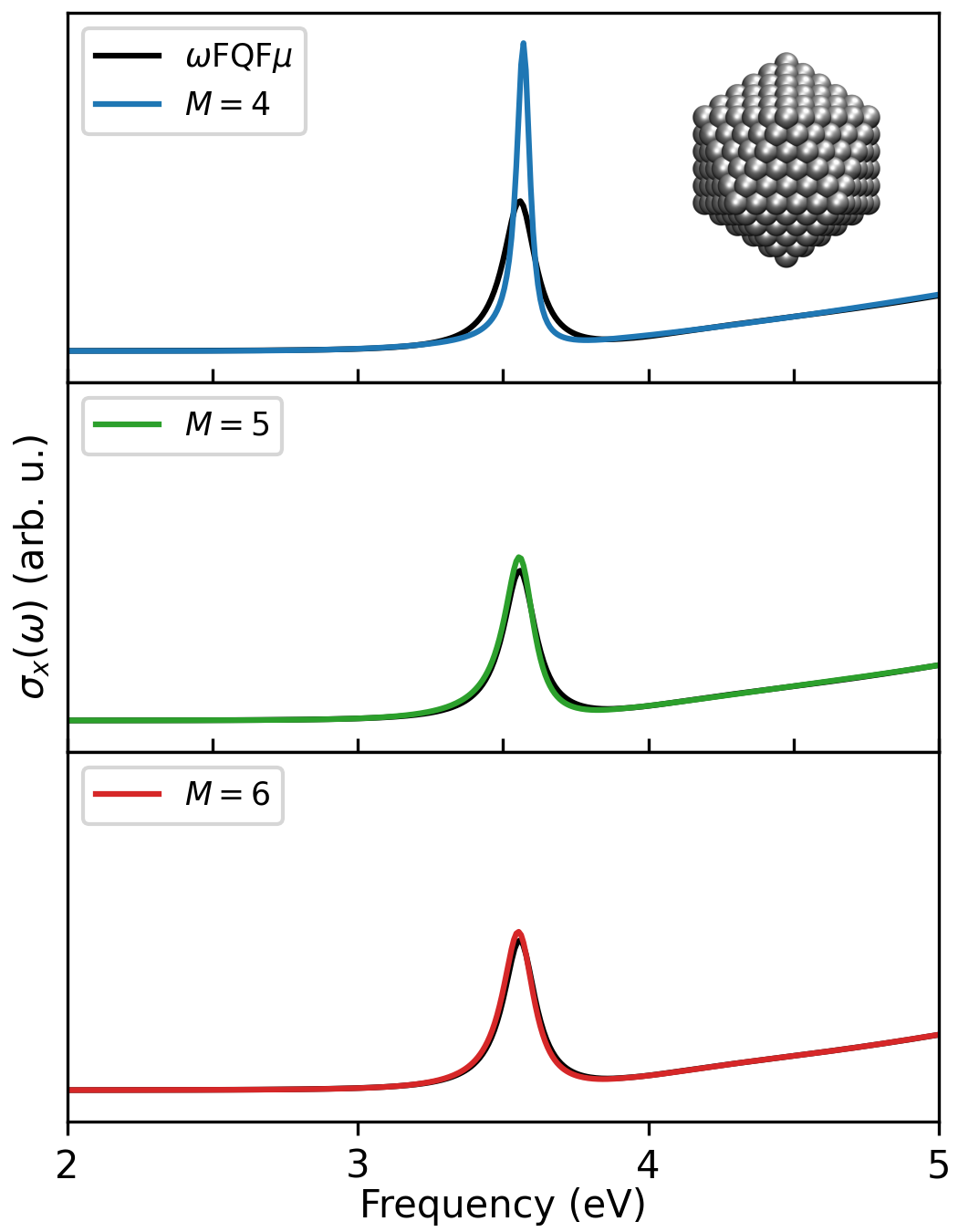}
    \caption{(left) Fitting of the experimental interband polarizability $\alpha^{\text{IB}}(\omega)$ by using $M = 4, 5, 6$ DL oscillators (see \cref{eq:alpha-ib}). (right) RT-\wfqfmu absorption cross section using $\alpha^{\text{IB}}$ as fitted using $M = 4, 5, 6$ DL oscillators. \wfqfmu reference spectrum as calculated by using the experimental $\alpha^{\text{IB}}$ is also reported for comparison's sake.}
    \label{fig:ag-fitting}
\end{figure}

Let us now consider the intricate case of $d$-metal nanostructures, for which a proper description of interband transitions is crucial to reproduce the correct experimental optical properties.\cite{giovannini2022we} As explained in \cref{sec:theory_rt}, in order to study the dynamics of \wfqfmu plasmonic response, we approximate the interband polarizability $\alpha^{\text{IB}}(\omega)$ as a sum of $M$ DL oscillators. The number $M$ of DL oscillators is chosen to ensure the accurate reproduction of the interband polarizability. To select the minimum number of DL required for this purpose, in \cref{fig:ag-fitting} (left) we report $\alpha^{IB}_{fit}$ as fitted by exploiting 4, 5, and 6 DL oscillators  For all cases, the reproduction of the experimental polarizability is particularly satisfactory, especially for frequencies larger than 4 eV. Some larger discrepancies are instead observed for smaller frequencies (1-4 eV), because $\alpha^{\text{IB}}(\omega)$ imaginary part is close to zero, and the DL functional form is not ideal in this case. The numerical fitting values are reported in table S1 in \sm. Note however that in such a region the optical response is dominated by intraband transitions, thus the effect of interband transitions is expected to be small. It is also worth pointing out that some amplitudes $A_p$ are negative (as also obtained in ref. \citenum{dall2020real}), however, the overall sign of the interband polarizability is positive.

To further evaluate the accuracy of the fitting procedure, in \cref{fig:ag-fitting} (right), the absorption cross-section of an Ih Ag NP constituted of 561 atoms (Ag$_{561}$, radius  $\sim$ 1.4 nm) is computed at the RT-\wfqfmu level ($\Delta t=1$ as, $T = 500 $ fs) by exploiting $\alpha^{\text{IB}}(\omega)$ fitted using $M=4, 5, 6$ DL oscillators. \wfqfmu spectrum calculated in the frequency domain by using the experimental interband polarizability is taken as a reference (black line in \cref{fig:ag-fitting}, right). The results clearly show that an accurate reproduction of the \wfqfmu spectrum is achieved when $M > 4$ DL oscillators are exploited in the fitting procedure. In fact, $M = 4$ (top panel) provides an overall good description of the \wfqfmu spectrum, however, the maximum intensity of the plasmon peak (at about $3.5$ eV) and its width are wrongly predicted (the intensity is almost twice the reference, the width is smaller). Indeed, our results remark on the crucial importance of a proper modeling of the interband transitions to reproduce the optical properties of noble metal NPs, highlighting how small differences in the interband polarizability can drastically affect the optical response. As a final comment, it is worth pointing out that the outlined discrepancies are associated with the diverse description of the interband polarizability. This is demonstrated by the perfect agreement between \wfqfmu and RT-\wfqfmu spectra computed by using the same permittivity function (see fig. S1 in the \sm).

\begin{figure}
    \centering
    \includegraphics[scale=1]{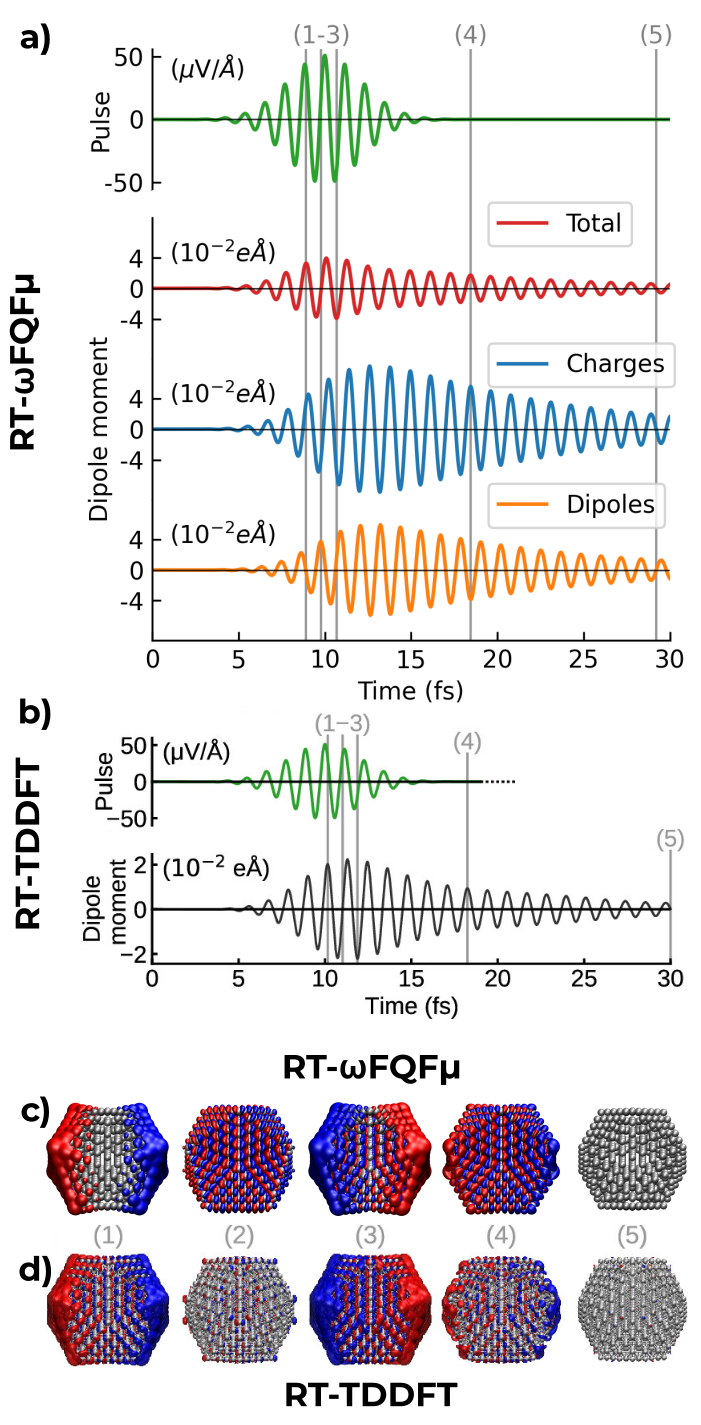} \\
    \caption{(a-b) RT-\wfqfmu (a) and RT-TDDFT\cite{rossi2020hot} (b) time evolution of the plasmon excitation of \ce{Ag561} Ih NP excited using a monochromatic Gaussian pulse (top panel). In panel a, the total RT-\wfqfmu response (red) is also decomposed in charge (blue) and dipole (orange) components. (c-d) RT-\wfqfmu (c) and RT-TDDFT\cite{rossi2020hot} (d) plasmon densities at specific times (1-5) as depicted in panels a-b, respectively. The isovalue is set to 0.035. RT-TDDFT data in panels b and d are adapted with permission from Ref. \citenum{rossi2020hot}.}
    \label{fig:ag-pulses}
\end{figure}

To further demonstrate the reliability and robustness of our newly developed RT-\wfqfmu approach, we simulate the time evolution of the plasmonic excitation of Ag$_{561}$ Ih NP under the effect of the following ultrafast Gaussian light pulse: 
\begin{equation}\label{eq:gaussian-pulse}
\mathbf{E}^{ext}(t) = E_0 \cos(\omega_0 (t-t_0))e^{-(t-t_0)^2/{(2\sigma^2)}}\hat{\mathbf{x}}
\end{equation}
where $E_0 = 51\, \mu V/$\AA\ and $\sigma = 2.121$ fs. $\omega_0$ is set to the PRF ($3.5$ eV). The parameters are chosen to match the \emph{ab initio} RT-TDDFT study performed by Rossi et al.\cite{rossi2020hot} on the same structure. Their \emph{ab initio} results are thus taken as a reference for challenging RT-\wfqfmu. A graphical depiction of the exploited impulse is given in \cref{fig:ag-pulses}, where the computed RT-\wfqfmu dipole as a function of time is also reported.
The light pulse triggers a plasmonic response by generating a pronounced dipole moment within the system, which arises after a few femtoseconds (see \cref{fig:ag-pulses}a). As the system evolves in time, specifically when $t$ approaches $13$ fs, the system undergoes a dephasing process. The coherence of the plasmon oscillations deteriorates, leading to a gradual decay of the dipole moment. The picture outlined by RT-\wfqfmu perfectly matches the reference \emph{ab initio} RT-TDDFT time evolution (see \cref{fig:ag-pulses}b). The agreement between the classical and quantum methods is excellent, not only qualitatively, but also quantitatively. In fact, only minor discrepancies are observed. The time delay of the plasmonic response for RT-\wfqfmu is smaller than RT-TDDFT as highlighted by vertical 1-3 bars in \cref{fig:ag-pulses}a-b. In addition, RT-\wfqfmu predicts a larger induced dipole moment, but of the same order of magnitude ($\sim 0.035$ e\AA~ vs. $\sim 0.022$ e\AA). This result is particularly remarkable, especially considering the classical nature of RT-\wfqfmu and the associated low computational cost.

To quantify the effective decay timing $\tau$ associated with the dephasing mechanisms, we fit the RT-\wfqfmu and RT-TDDFT curves by using $e^{-t/\tau}$ functional form. The resulting RT-\wfqfmu $\tau$ is $9.6$ fs, showing a typical localized surface plasmon decay time. This is in very good agreement with the reference \emph{ab initio} fitted $\tau$ value ($7.8$ fs). Therefore, RT-\wfqfmu can be as accurate as RT-TDDFT in describing the plasmonic time evolution and decay, even for nanostructures below the quantum size limit, where it is commonly accepted that an explicit quantum description is necessary. The small numerical discrepancy can be due to the fact that RT-\wfqfmu does not account for electron-surface scattering damping mechanisms\cite{giovannini2022we,khurgin2020generating,khurgin2017landau} thus modifying the final numerical value of $\tau$.
 
The fitted $\tau$ substantially differs from the Drude scattering time exploited to model intraband mechanisms in our RT-\wfqfmu calculations ($\sim 39$ fs). Such a finding suggests that interband effects play a non-trivial role in determining the time evolution of the response of the plasmonic decay. This is expected since interband transitions determine the optical response of noble nanostructures as demonstrated in \cref{fig:ag-fitting}. In fact, the interband relaxation in Ag nanostructures is governed by the inverse of $\gamma_p$ in \cref{eq:wfqfmu-rt-dipoles} (see also table S1 in the \sm). For $M = 6$ DL oscillators, the scattering times ($1/\gamma_p$) range from $4.5$ as to $30.5$ as. Therefore, intra and interband decay mechanisms non-trivially interplay, resulting in the fitted $\tau$ value. To deepen on this point, in \cref{fig:ag-pulses}a, the RT-\wfqfmu time evolution of the total dipole moment is decomposed in its charge and dipole contributions. In fact, our classical approach allows for dissecting the contributions from delocalized and $d$ electrons. This is possible thanks to the formulation of RT-\wfqfmu in terms of charges and dipoles with a clear physical meaning being directly associated with intra and interband mechanisms, respectively.\cite{giovannini2022we} \Cref{fig:ag-pulses}a shows that charges and dipole dynamics are inherently connected, generally oscillating in counter phase for the whole duration of the plasmonic excitation (especially for $t > 13$ fs). Also, note that both charge and dipole decay with the same effective time, although the time scattering related to the two underlying mechanisms is defined in two different timescales. This is not surprising considering that charges and dipoles act as coupled oscillators in our model (see \cref{eq:wfqfmu-rt-charges,eq:wfqfmu-rt-dipoles}). However, their dynamics non-trivially couple, thus resulting in the total dipole moment time evolution which has been previously discussed.  

To further investigate this point, we analyze the time evolution of the plasmonic response by plotting the plasmon densities at relevant times (see 1-5, in \cref{fig:ag-pulses}a,b), which are given in \cref{fig:ag-pulses}c. The RT-TDDFT plasmon densities, reproduced from Ref. \citenum{rossi2020hot}, are given in \cref{fig:ag-pulses}d as a reference. The plasmonic excitation is characterized by electron density oscillations, which can be dissected into two main components. The first component is a surface-to-surface element, predominantly associated with delocalized valence electrons and intraband mechanisms. These electrons exhibit collective behavior and mainly determine the overall electron density oscillations. The second component arises from atom-localized contributions, which are linked to the screening effect due to interband transitions originating from the $d$-band. The plasmon decay (plots 4-5 in \cref{fig:ag-pulses}c) is directly correlated with the diminishing surface-to-surface electron density oscillations, marking the end of the collective plasmonic activity. The intricate interplay between the delocalized valence electrons and the localized $d$-band electrons therefore forms the basis of the observed electron density oscillations. By comparing RT-\wfqfmu and RT-TDDFT, we note that such an interplay is correctly reproduced by our classical RT-\wfqfmu approach, thanks to the proper inclusion of intraband and interband decay mechanisms. 


%
%
%
%
%
%
%
%
%
%
%

\section{Summary and Conclusions}

In this work we have presented two novel approaches, namely RT-\wfq and RT-\wfqfmu, to describe the time evolution of plasmonic excitations in simple and noble metal nanoparticles. 
The developed methods are the real-time extension of \wfq and \wfqfmu, which have been previously developed in the frequency domain. Such models provide an accurate fully atomistic modeling of plasmonic nanoparticles, by properly describing intra and interband effects and also tunneling mechanisms. Our real-time extension can handle short impulses (in the as/fs timescale) and CW within the same theoretical formulation.

RT-\wfq and RT-\wfqfmu are first numerically validated taking the frequency-domain counterparts' results as reference. Then the models have been tested against two different challenging nanostructures: a sodium dimer characterized by a single atom junction, and an Ih Ag NP. To show the flexibility of our implementation, the first system is studied under CW illumination, while a Gaussian pulse is used to excite the Ag NP. As a result, while in the first case, after a transient, the total dipole moment of the nanostructure oscillates in phase with the external CW field, Ag NP total dipole moment is first excited and then fast decays in the fs timescale. In the latter, we directly compared with \emph{ab initio} RT-TDDFT, demonstrating a qualitative and even quantitative agreement between our classical and the quantum approach. As a consequence, our approach can serve as a powerful approach for simulating the plasmonic response dynamics even for NP size below the quantum size limit. We note that we can further increase the agreement between our approach and full quantum treatments by including electron surface scattering damping decay mechanisms, which can serve as an additional decay channel.\cite{khurgin2020generating}

The development of RT-\wfq and RT-\wfqfmu paves the way for the investigation of the decoherence of the plasmonic response in nanostructures of sizes considerably larger than those that can be treated at the purely quantum level, without losing accuracy. Also, the model has the potential to be coupled to a QM description of molecular systems adsorbed on metal nanostructures,\cite{huang2024time} as previously done in the frequency domain\cite{lafiosca2023qm,illobre2024multiscale} or by using diverse electrodynamical approaches.\cite{pipolo2016real,dall2020real,marsili2022electronic,coccia2020hybrid,dall2024peeking} Such an extension will be the topic of future papers. 

\section*{Acknowledgments}

We thank Giulia Dall'Osto for the discussions on the dielectric function fitting. We gratefully acknowledge the Center for High-Performance Computing (CHPC) at SNS for providing the computational infrastructure.

\subsection*{Funding Sources}

This work has received funding from the European Research Council (ERC) under the European Union’s Horizon 2020 research and innovation programme (grant agreement No. 818064). SC thanks the Horizon 2020 EU grant ProID (grant agreement No. 964363) for funding.

\begin{suppinfo}
Numerical values of $A_p,\omega_p$, and $\gamma_p$ for $M = 4, 5, 6$ DLs. Graphical depiction of plasmon time evolution for B, C, D, and E peaks depicted in \cref{fig:na380-dynamics}.
\end{suppinfo}

\bibliography{biblio}

\end{document}